\documentstyle[epsf,12pt]{article}
\def\comment#1{}

\title{\LARGE \bf High-Performance Special-Purpose Computers in Science}
\date{}
\author{\large Toshiyuki Fukushige$^*$\\
\large Piet Hut$^+$\\
\large Junichiro Makino$^*$\\
\large $*$ Department of Systems Sciences, College of Arts and Sciences,\\
\large University of Tokyo\\
\large $+$Institute for Advanced Study
}

\begin{document}

{

\baselineskip 20 pt

\maketitle

}

The next decade will be an exciting time for computational physicists.
After 50 years of being forced to use standardized commercial
equipment, it will finally become relatively straightforward to adapt
one's computing tools to one's own needs.  The breakthrough that opens
this new era is the now wide-spread availability of {\it programmable
chips} that allow virtually every computational scientist to design
his or her own special-purpose computer.

\newpage

\subsection*{Towards Real Numerical Laboratories}

Unlike real laboratories, numerical laboratories have been constructed
almost exclusively from commercial products, which gave little
flexibility.  Starting in the late seventies, after the first
microchips became available, there have been some exceptions.
However, only the bravest souls dared to design their own equipment
(Bakker and Bruin 1988).

In those days, speeding up the most compute-intensive few lines of
FORTRAN code, in a large-scale simulation project, required
building a bulky piece of electronic hardware consisting of tens of
large circuit boards.  By the late eighties, things looked a lot
better already, since it had become possible to integrate such
circuits into a single custom chip.  However, the barrier, real or
perceived, against building your own hardware was still substantial.

By now, after another ten years, the barrier has almost disappeared.
With programmable chips, the question is not so much whether to make
one's hands dirty building a machine, but rather how to program:
whether to program the existing central processing unit of a
commercial machine, or whether to program a more generic set of
distributed processing units.  In both cases, the trick is to find the
best map between the scientific problem and the layout of the
computational hardware.  What favors the programmable option over the
use of standard CPUs are the facts that: 1) supercomputers, optimized
for scientific calculations, are rapidly disappearing, leaving us with
less-than-optimal vanilla-flavored computers; 2) generic CPU chips are
now becoming so complex that the overwhelming fraction of silicon real
estate is dedicated to the electronic equivalent of bureaucracy rather
than raw computing.

Of course, there is still one major drawback to the use of
programmable chips in computational science: habit.  It takes a while
for scientists to switch their approach to a problem, even when more
efficient methods have become available.  Therefore, in order to
prepare for the future, it is useful to look back at the past, to see
what has already been accomplished during the last ten years, using
special-purpose computers.  Anything that has been done in this area,
relying on specially designed chips, can in principle be done now with
programmable chips, at only a fraction of the effort involved.  Let us
focus on a specific case.

\subsection*{               A Case Study: The GRAPE Project}

Our GRAPE project, started 10 years ago, is one example in which
computational physicists developed special-purpose computers
successfully. Here, success means that the developed machine made it
possible to solve problems which were impossible to solve on
general-purpose computers.

One of these projects has resulted in the GRAPE (short for GRAvity
PipE) family of special-purpose hardware, designed and built by a
small group of astrophysicists at the University of Tokyo (Makino and
Taiji 1998). Like a graphics accelerator speeding up graphics
calculations on a workstation, without changing the software running
on that workstation, the GRAPE acts as a Newtonian force accelerator,
in the form of an attached piece of hardware.  In a large-scale
gravitational N-body calculation, almost all instructions of the
corresponding computer program are thus performed on a standard work
station, while only the gravitational force calculations, in innermost
loop, are replaced by a function call to the special-purpose hardware.

The GRAPE-4, which was completed in 1995 for the total budget of 240 M
JYE (around 2 M dollars), offered a peak speed of 1.08 Tflops. On
practical problems, a significant fraction of this speed can be
actually used.  For example, the Grape-4 developers have won the
Gordon Bell prize for high-performance computing for two years in a
row.  In 1995, the prize was awarded to Junichiro Makino and Makoto
Taiji for a sustained speed of 112 Gflops, achieved using one-sixth of
the full machine on a 128k particle simulation of the evolution of a
double black-hole system in the core of a galaxy.  The 1996 prize was
awarded to Toshiyuki Fukushige and Junichiro Makino for a 332 Gflops
simulation of the formation of a cold dark matter halo around a
galaxy, modeled using 768k particles on three-quarters of the full
machine.  The first general-purpose computer to offer a similar level
of the performance is the 9000-processor ASCI Red machine, with a
price tag around 50 M dollars, completed in late 1997.

In addition, more than 40 copies of small (5-30 Gflops) GRAPE-3 and
GRAPE-4 versions are now being used in a major astrophysical
institutes in many different countries.

In a year from now, the GRAPE-6 will become available, at a speed that
will be at least 100 times faster than that of the GRAPE-4.  In
addition, single board versions will become available (the `GRAPE-6
junior'), that can be purchased by individual astrophysicists, to run
at a speed of 500 Gflops, coupled to a normal workstation as a
front-end.  Such a single board will thus provide a speed-up of well
over a factor $1,000$ for a price comparable to that of the
workstation.

What is the main reason behind the success of the GRAPE?  From a
technological point of view, it is not overly difficult to design a
special-purpose computer with a cost-performance better than that of
commercially available general-purpose computers. The reason is simply
that an ever diminishing fraction of the available transistors in the
present-day microprocessors are actually used in arithmetic
operations. In contrast, in GRAPE systems essentially all available
transistors in a processor chip are used to implement arithmetic
units.

This main reason behind the optimal usage of silicon real estate in
the GRAPE is that the data flow in a GRAPE chip is fixed, while most
of the transistors in present-day microprocessors are utilize to
provide flexible data flow.  In other words, the key trick to
outperform general-purpose machines is to build a hardware accelerator
for a specialized function, and not to build a programmable
computer. To design a programmable computer is a difficult task, and
you have a very little chance (unless you are a Seymour Cray) to
outwit competent designers in large companies.  However, designing a
specialized piece of hardware for a single function is not something
computer architects do. So there is essentially no competition.

\subsection*{                      Having it all}

The interesting shift, alluded to in our opening statement, is that it
is no longer necessary to choose between programming a computer and
building a special-purpose computer.  With the availability of
increasingly efficient programmable chips, one can {\it program} an
off-the-shelf chip, such as a FPGA (Field-Programmable Gate Array)
chip, to {\it emulate a special-purpose chip}.  It is like having your
cake and eating it: you can emulate a fixed date flow.  Before you
program the chip, it is far more flexible than a standard CPU, in the
sense that you are not bound by a given instruction set, providing all
instructions yourself, from the bottom up.  And after having
programmed the chip, it has turned into a data flow machine, without
any need to decode additional information on the fly.

\begin{figure}
\begin{center}
\leavevmode
\epsfxsize 10 cm
\epsffile{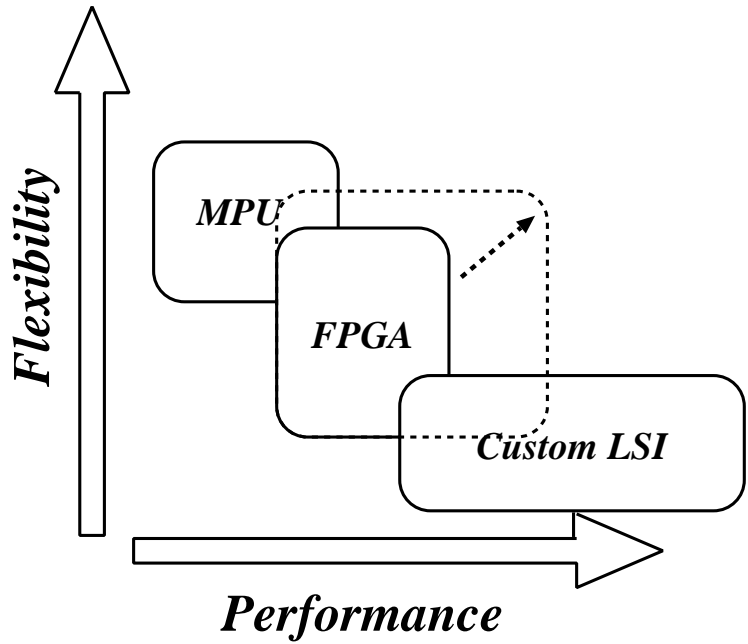}
\caption{Relation between programmable microprocessors (MPU), FPGA, and custom
LSIs. Future FPGA will offer more flexibility and higher performance.}
\end{center}
\end{figure}

Of course, there is a drawback to any new development.  With current
FPGAs, flexibility comes with a cost: more than 90\% of the silicon
resource is used to provide programmability. Even so, custom computing
machines based on FPGA have become a viable alternative for both
general-purpose computers and specialized hardwares, given the fact
that standard CPUs tend to have lower and lower efficiency as well,
while their complexity keeps increasing.
	
We have developed a small system with two FPGA chips to evaluate the
potential of FPGA technology.  The current FPGAs turned out to be
large enough to house a complete GRAPE-3 chip (110K transistors). The
chips available in next year will be large enough to house a GRAPE-4
chip and the effective performance would exceed 1 Gflops per chip.

\subsection*{                     Outlook}

To summarize, FPGAs offer the possibility of combining the flexibility
of conventional programmable computer and the high throughput of
special-purpose hardware.  To play the devil's advocate, one might
argue that FPGAs could combine the difficulty of design of
special-purpose hardware and the low efficiency of the programmable
computer.  To be honest, at present there still is that danger.  To
continue the devil's advocate argument: implementing a function onto an
FPGA is analogous to programming a universal Turing machine, in the
sense that it offers the maximum flexibility at the lowest level.
Clearly, a more sensible design methodology is necessary.  On the
bright side, rapid advances are being made in the development of
higher level tools for implementing algorithms on FPGAs.  And the more
the parallelization bottleneck, using general-purpose computers, will
be felt, the larger the incentive will become to switch to a use of
programmable chips.

Thus FPGAs are not a universal solution for all problems in
computational physics.  A full-custom chip offers clear advantages
when its high initial development cost can be amortized by mass
production.  General-purpose computers are still better in developing
sophisticated algorithms, and experimenting with them.  But the use of
FPGAs can be expected to increase rapidly, in computational science,
for a wide range of problems, that are too complex to `put in stone'
in the form of a special-purpose chip, but not too complex to program
onto a FPGA.  In that way, problems which cannot be solved in a
practical time on programmable computers, do not have to be shelved
until commercial computers catch up and deliver the required speed.
As the example of the GRAPE has shown us, even a small group of
computational scientists can solve their particular computational
problems years ahead of (the commercial) schedule.

\subsection*{References}

Bakker A. F. and Bruin C. (1988) Design and implementation of the
Delft molecular-dynamics processor.  In Alder B. J. (ed) Special
Purpose Computers, pages 183--232.  (Academic Press, San Diego)

\smallskip

\noindent
Makino J. and Taiji M. (1998) Special Purpose Computers for Scientific
Simulations -- The GRAPE systems.  (John Wiley and Sons, Chichester)

\end{document}